# Spontaneous fluctuations of transition dipole moment orientation in OLED triplet emitters

*Florian Steiner, Sebastian Bange, Jan Vogelsang, and John M. Lupton*[*]

Institut für Experimentelle und Angewandte Physik, Universität Regensburg,

Universitätsstrasse 31, 93053 Regensburg, Germany

**Corresponding Author**

*john.lupton@physik.uni-regensburg.de

## ABSTRACT

The efficiency of an organic light-emitting diode (OLED) depends on the microscopic orientation of transition dipole moments of the molecular emitters. The most effective materials used for light generation have threefold symmetry, which prohibit *a priori* determination of dipole orientation due to the degeneracy of the fundamental transition. Single-molecule spectroscopy reveals that the model triplet emitter *tris*(2-phenylisoquinoline)iridium(III) (Ir(piq)$_3$) does not behave as a linear dipole, radiating with lower polarization anisotropy than expected. Spontaneous symmetry breaking occurs in the excited state, leading to a random selection of one of the three ligands to form a charge transfer state with the metal. This non-deterministic localization is revealed in switching of the degree of linear polarization of phosphorescence. Polarization scrambling likely raises out-coupling efficiency and should be taken into account when deriving molecular orientation of the guest emitter within the OLED host from ensemble angular emission profiles.

## TOC GRAPHICS

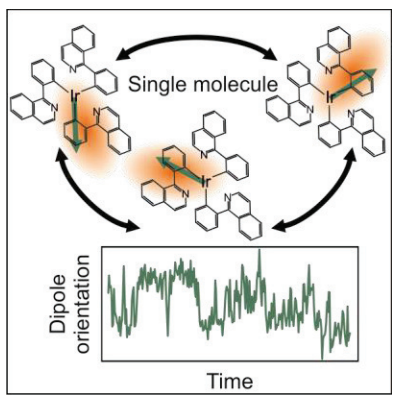

The operating principle of organic light-emitting diodes (OLEDs) is based on the injection of electrons and holes from opposite electrodes. Carriers bind electrostatically, ultimately recombining. Since spin-orbit coupling is weak in materials of low atomic order number, the electronic spin remains a good quantum number, defining dipole selection rules. Up to three quarters of recombination events occur within the triplet manifold of the molecular material, which is generally non-emissive, limiting overall device quantum efficiency. Almost 40 years passed from the first demonstration of OLEDs[1,2] to the introduction of a simple fix to the spin problem: inclusion of metal-organic complexes in the emissive layer.[3] These materials show strong triplet emission in the form of phosphorescence. Typical triplet emitters are based on metal-ligand charge transfer complexes,[3,4,5] such as metalloporphyrins or iridium bispyridines, which must be embedded in a host material to prevent concentration quenching. These materials are all characterized by non-trivial symmetry and resulting level degeneracies. The commonly used *tris*(2-phenylisoquinoline)iridium(III) (Ir(piq)$_3$), for example, has threefold (C$_3$) symmetry which leads to different excited triplet (T$_1$) to singlet ground state (S$_0$) transitions (T$_1$ $\rightarrow$ S$_0$). It was shown for such C$_3$-symmetric molecules by Jansson *et al.* that the T$_1$ $\rightarrow$ S$_0$ transition has charge transfer character and involves one of the three ligands with the central metal atom.[6] Given that in an OLED, light emission is of foremost interest, it is a surprisingly non-trivial question to consider the spatial orientation of the underlying transition dipole moment (TDM) in such a degenerate system. Conventional luminescence-based methods to determining polarization anisotropy[7,8] fail since rotational diffusion times in solution are far shorter than the radiative lifetime.

The orientation of the dipole of the emitting species determines the overall device efficiency.[9,10,11] As sketched in Fig. 1a), dipole orientation along the normal of the emitting layer will strongly reduce radiative coupling into the far field and increase in-plane wave guide losses. While light scattering promoted by lateral structuring can reduce such wave-guide losses,[12] it would be desirable to control dipole orientation within the emitting layer so as to maximize out-coupling to start with.[13] With OLED quantum efficiencies now approaching 40 % in the red spectral range, it is becoming apparent that some combinations of organometallic host-guest OLED materials appear to promote anisotropic molecular arrangements to yield preferential orientation of the TDM within the OLED plane.[13,14,15,16] Such anisotropy can be visualized by the deviation from a Lambertian angular dispersion of the emission spectrum.[17,18] These ensemble measurements, however, only place a lower limit on the anisotropy of the emitter TDM and therefore actually understate the degree of spontaneous ordering in the host-guest system. From ensemble ultrafast spectroscopy, localization of the photoexcited singlet metal-ligand charge transfer state is known to occur within a few hundred femtoseconds.[8,19] But which of the three ligands is responsible for ultimate emission from the triplet state?

Here, we demonstrate how polarization-resolved single-molecule spectroscopy[20,21,22,23,24,25] can be applied to a common organometallic triplet emitter, *tris*(1-phenylisoquinoline)iridium(III) (Ir(piq)₃) to reveal spontaneous symmetry breaking in the excited state.[26] The effect becomes apparent in temporal fluctuations of the TDM vector as reported by the linear dichroism in single-molecule phosphorescence. Such fluctuations are likely to be a universal feature of molecules of higher symmetry and will influence OLED efficiency since the orientation of the TDM controls out-coupling efficiency. Given the possibility of spontaneous alignment of the guest within the host,[17,18] our results confirm why it can be desirable to break the common threefold molecular symmetry *a priori*, lifting the degeneracy.[13]

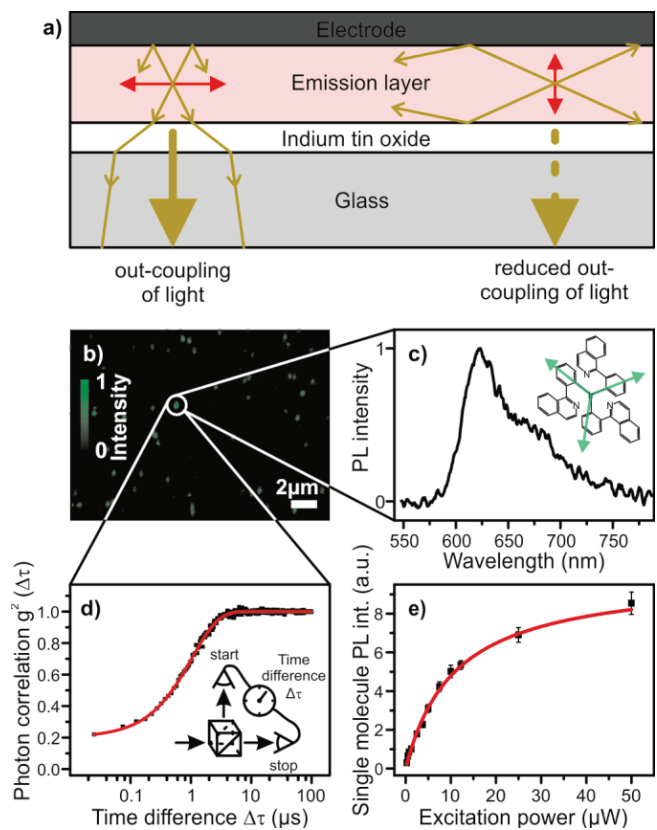

**Figure 1. Single-molecule spectroscopy of triplet emitters in OLEDs and the effect of dipole orientation on light out-coupling.** a) The out-coupling efficiency is controlled by

dipole orientation (red arrow). Radiation polarized in the plane of the emission layer is preferentially trapped by wave-guiding effects. b) A scanning confocal single-molecule microscope image of Ir(piq)$_3$ under excitation at 485 nm (chemical structure shown as inset in c). The single molecules all show comparable brightness and allow extraction of the PL spectrum (c). d) Photon correlation spectroscopy of the single-molecule emission. The phosphorescence is passed through a beam splitter and the coincidence rate between two photodiodes is measured. The correlation shows a photon antibunching dip at short times, implying that only one photon passes through the beam splitter at a given time. The red line is a single exponential fit to the correlation which is determined by the apparent triplet lifetime. e) Single-molecule PL intensity as a function of excitation power averaged for approximately 300 molecules for each excitation intensity. The red line shows the fit with a saturation function as described in the *Supporting Information*.

Single Ir(piq)$_3$ molecules (structure inset in Fig. 1c) can be imaged, at room temperature under a nitrogen atmosphere, by serial dilution down to picomolar concentration and dispersion in a poly(methyl-methacrylate) (PMMA) matrix (see *Supporting Information* for details on sample preparation). Given the success of single-molecule techniques in applications to a wide range of problems from the life- to the materials sciences, there have been surprisingly few attempts to study individual organometallic triplet emitters.[27,28,29,30,31,32] The obvious limitation is that photon emission rates in phosphorescence are fundamentally limited by the triplet excited state lifetime, which is of the order of 1 μs, two to three orders of magnitude longer than for typical singlet fluorophores. Figure 1b) shows a scanning confocal single-molecule microscope image under excitation with a continuous-wave laser at 485 nm (see *Supporting Information* for details on methods). Discrete spots of approximately uniform brightness are seen which show the characteristic Ir(piq)$_3$ photoluminescence (PL) spectrum in panel c). It is straightforward to demonstrate that each spot corresponds to a single phosphorescent molecule by considering the temporal correlation in photons emitted. The photons from a single diffraction-limited spot are passed through a 50/50 beam splitter and recorded with two different photodetectors in a Hanbury-Brown and Twiss arrangement (sketch inset in panel d). Since a single photon cannot be recorded by both detectors simultaneously, for times shorter than the excited state lifetime a reduction in the correlation coefficient is observed in panel d). This photon antibunching is generally only considered for short-lived (singlet) excited states in atoms[33,34] and molecules,[35,36] but actually works surprisingly well for triplet emitters[31] since the arrival times of the detected photons provide an implicit mechanism to discriminate from short-lived background fluorescence or impurities. The second-order correlation function g$^2$ as a function of delay time $\Delta\tau$ in panel d) is described by a simple exponential function (red line, see *Supporting Information*), with an apparent lifetime determined by the triplet excitation and decay rates. Since the single-molecule PL intensity depends on both the excitation and the emission rate, the brightness saturates as the excitation rate approaches the emission rate. This saturation effect, which only becomes discernible on the single-molecule level, is shown in Fig. 1e) as a function of excitation power for an average of ~300 single molecules for each measured excitation intensity, and is accurately described by a saturation curve (solid line, see *Supporting Information* for details).

The orientations of the three degenerate main triplet TDMs of Ir(piq)$_3$ involve charge transfer from one of the ligands to the metal center. These TDMs have been computed previously[6,13] and are sketched in Fig. 1c). A single-molecule experiment allows determination both of the absolute orientation of the TDM with respect to the observer's reference frame, but can also unveil possible fluctuations in orientation. 3D mapping of TDM orientation requires high photon count rates,[24,37] which are not compatible with triplet emitters. The *average* anisotropy of the TDM of many single molecules along with relative changes in orientation over time within one single molecule can, however, easily be analyzed by passing the PL through a polarizing beam splitter and recording two mutually perpendicular polarization components on separate detectors, as sketched in Fig. 2a). Such an experiment provides a metric for polarization anisotropy: linear dichroism, defined as the difference-sum-ratio between the two polarization components as stated in the figure. The absolute value of this linear dichroism is only meaningful in a statistical analysis of randomly oriented single molecules, which all individually exhibit photon antibunching. Fig. 2b) shows a histogram of linear dichroism values for 460 single molecules. In contrast to molecules with a single TDM (i.e. fully anisotropic),[38] no linear dichroism values of $\pm 1$ are seen, implying that the luminescence is never perfectly linearly polarized. This observation is rather unusual for such a small single molecule. To analyze the distribution of linear dichroism values, we calculated the expected linear dichroism histogram for an ensemble of randomly oriented single dipoles in either three-dimensional space or in a the two-dimensional sample plane, taking into account the collection angle (numerical aperture) of the microscope (see *Supporting Information* for details).[38] The results are shown as orange and blue dots in Fig. 2b), respectively, which display substantial weighting to higher absolute linear dichroism values, in marked contrast to the measurement. We conclude that Ir(piq)$_3$ does not behave as a single dipole emitter on the timescales of this experiment (on average 100 seconds integration time per molecule).

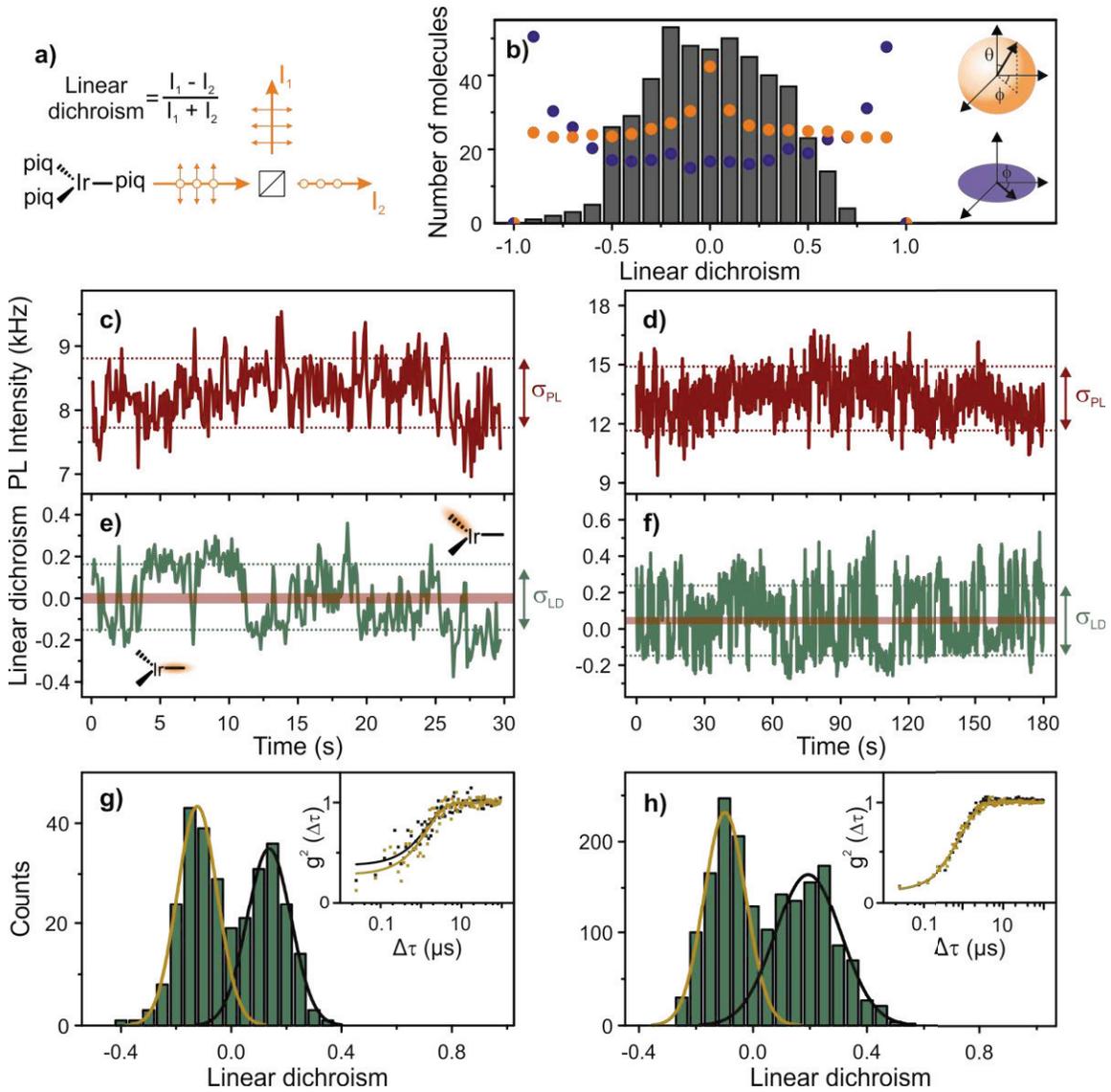

**Figure 2. Spontaneous fluctuations in single-molecule transition dipole moment (TDM) orientation.** a) The light is passed through a polarizing beam splitter and detected by two photodiodes. The linear dichroism is defined as the difference-sum-ratio of the two polarization planes. b) Histogram of linear dichroism values for 460 single

molecules. The dots show expected distributions for transition dipoles oriented randomly in three-dimensional space (orange) or in two dimensions (blue), taking into account the collection aperture of the microscope. c), d) Two representative single-molecule PL intensity traces and corresponding linear dichroism values (e, f). The noise on the PL is defined by the standard deviation in the intensity as $\sigma_{PL}$ as indicated by the dotted lines. This value is used to compute the expected noise in the linear dichroism measurement (indicated by the red bars). The measured noise in the linear dichroism is defined as $\sigma_{LD}$. g), h) Histograms of linear dichroism values for the two single molecules, showing two preferential orientations of TDM. These two orientations correspond to the same excited state lifetime, shown in the photon cross correlation $g^2(\Delta\tau)$ of the emission for photons selected by linear dichroism value (marked black and yellow).

In addition to this statistical analysis, we consider the possibility of dynamics in the emission polarization for two single molecules. Panels c) and d) plot the total PL intensity, the sum of both polarization channels, given in photon count rate. The single-molecule brightness shows a certain level of fluctuations which is due both to the detector shot noise and intrinsic dynamics of the molecule. This overall noise in the PL signal is quantified by the standard deviation from the mean and marked in the figure as $\sigma_{PL}$. The corresponding linear dichroism, binned in 100 ms intervals, is shown in panels e) and f) for the two molecules. Both cases display clear discrete jumps in polarization, which imply a switching of TDM orientation. In addition, panel f) also shows fluctuations at higher frequency. The polarization fluctuations can be quantified by defining the standard deviation of the linear dichroism trace $\sigma_{LD}$ from the measurement. For comparison, we mark the noise in linear dichroism calculated from the number of detected photons in each detection channel, taking into account error propagation, as red bars in panels e), f). The measured noise is much greater than the calculated noise, implying that it does not originate primarily from the shot noise in the PL. The linear dichroism fluctuations can be analyzed by plotting the values in histograms in panels g), h). In both cases two distinct peaks are found, which are accurately described by Gaussians: on average, two orientations of the TDM can be resolved. However, since there is more high-frequency noise in the right-hand example, the two Gaussians are broadened and overlap more strongly in this case. We conclude that fluctuations in TDM orientation appear not only as discrete jumps but also as an overall increase in scatter of linear dichroism values, described by the standard deviation $\sigma_{LD}$.

It is important to confirm that the two measured TDM orientations of one and the same single molecule (Fig. 1g,h) correspond to the same excited state species. We therefore compare the photon correlation (as described in Fig. 1d), but perform a selection in terms of the value of linear dichroism. The two subsets marked yellow and black in the figure exhibit exactly the same photon correlation function, implying that the distinct values in linear dichroism arise from a molecular excited state with identical lifetime, even though the lifetime scatters between different single molecules. This agreement is expected from the rotational symmetry of the molecule: switching TDM orientation will not affect TDM magnitude.

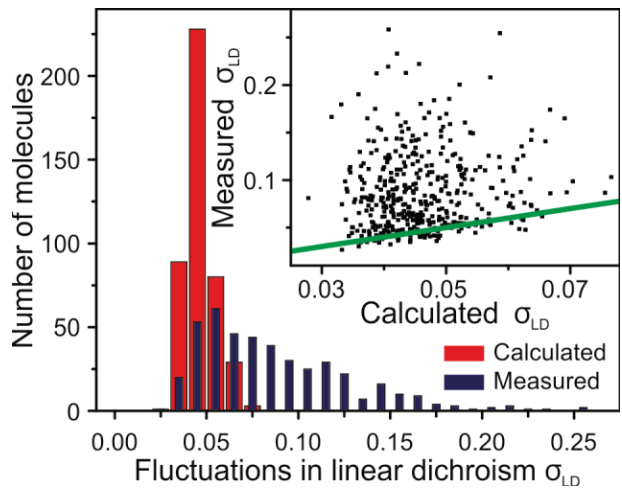

**Figure 3. Non-deterministic TDM orientation in 460 single Ir(piq)₃ molecules revealed by the comparison of measured to computed noise in linear dichroism $\sigma_{LD}$.** The calculated $\sigma_{LD}$ values are computed based on the observed noise in the PL intensity, $\sigma_{PL}$. Inset: the measured $\sigma_{LD}$ values mostly lie significantly above the calculated values. The green line shows the limit where measured $\sigma_{LD}$ equals calculated $\sigma_{LD}$.

The detected jump in TDM depends sensitively on the absolute orientation of the molecule with respect to the plane of the polarizing beam splitter. In many cases, discrete polarization jumps will be masked by the noise on the linear dichroism measurement, so that discrete peaks in the histogram of linear dichroism values of a single molecule are obscured. Nevertheless, the noise $\sigma_{LD}$ provides a direct metric for TDM fluctuations since it can be compared to the expected noise calculated from the PL intensities on each polarization channel: the two observables are obtained from the same measurement. Figure 3 compares the calculated to the measured values of $\sigma_{LD}$ for 460 single molecules. Based on the PL noise, the maximal fluctuation in linear dichroism expected is 0.075, whereas values of up to 0.25 are found experimentally. In virtually all cases the measured $\sigma_{LD}$ is significantly larger than the calculated value based on the PL noise, as shown in the inset of Figure 3 where the measured $\sigma_{LD}$ is plotted against the calculated $\sigma_{LD}$. The green line shows the limit where both values are the same.

The TDM orientation of a common OLED triplet emitter fluctuates in time. This unusual effect is not found in conventional chromophores with well-defined TDM orientations such as typical fluorescent dye molecules used in biophysical applications. The fluctuations arise from the molecular symmetry which determines the degeneracy in transitions to the ground state. We note that limited emphasis should be attached to the absolute measurements of linear dichroism (e.g. in Fig. 2b) since these values may be reduced in magnitude if rapid fluctuations occur between the two TDM orientations. These fluctuations clearly explain the marked deviation between measured and computed TDM orientation distribution in Fig. 2b). It was recently shown that a single ring-shaped (singlet-emitting) fluorophore displays almost completely unpolarized PL due to non-deterministic switching between TDM orientations following every excitation event.[39] Such a system allowed for the differentiation between spontaneous and photoinduced symmetry breaking.[39] If such fluctuation also occurs for emitters embedded within the OLED host matrix, measurements of ensemble TDM anisotropy in the film based on assessing the deviation from a Lambertian emitter dispersion profile[18] will actually *underestimate* the degree of molecular ordering. Our experiments show that the TDM orientation can fluctuate spontaneously and is likely to be very sensitive to the surrounding environment. Anisotropy in the ensemble could then occur due to pinning of the TDM orientation, which has been speculated to arise from self-organization of the host matrix due to interactions with the polar device substrate.[13]

ASSOCIATED CONTENT

**Supporting Information.** Details of sample preparation, experimental setup, further analysis and simulations. This material is available free of charge via the Internet at http://pubs.acs.org.

AUTHOR INFORMATION

**Corresponding Author**

*Email: john.lupton@physik.uni-regensburg.de

**Notes**

The authors declare no competing financial interests.

ACKNOWLEDGEMENT

The authors thank the ERC for financial support through the Starting Grant MolMesON (#305070) and the DFG-GRK 1570. JML is indebted to the David and Lucille Packard Foundation for funding a Fellowship.

REFERENCES

1    Pope, M., Magnante, P. & Kallmann, H. P. Electroluminescence in organic crystals. *J. Chem. Phys.* **38**, 2042-2045 (1963).

2    Helfrich, W. & Schneider, W. Recombination radiation in anthracene crystals. *Phys. Rev. Lett.* **14**, 229-232 (1965).

3    Baldo, M. A., Thompson, M. E. & Forrest, S. R. Phosphorescent materials for application to organic light emitting devices. *Pure Appl. Chem.* **71**, 2095-2106 (1999).

4    Rausch, A. F., Homeier, H. H. H. & Yersin, H. in *Photophysics of Organometallics* Vol. 29 *Topics in Organometallic Chemistry* (ed A. J. Lees) 193-235 (2010).

5    Meyer, T. J. Photochemistry of metal coordination-complexes - metal to ligand charge-transfer excited-states. *Pure Appl. Chem.* **58**, 1193-1206 (1986).

6    Jansson, E., Minaev, B., Schrader, S. & Agren, H. Time-dependent density functional calculations of phosphorescence parameters for fac-tris(2-phenylpyridine) iridium. *Chem. Phys.* **333**, 157-167 (2007).

7    Hedley, G. J., Ruseckas, A. & Samuel, I. D. W. Ultrafast luminescence in Ir(ppy)(3). *Chem. Phys. Lett.* **450**, 292-296 (2008).

8    Yeh, A. T., Shank, C. V. & McCusker, J. K. Ultrafast electron localization dynamics following photo-induced charge transfer. *Science* **289**, 935-938 (2000).

9    Wasey, J. A. E., Safonov, A., Samuel, I. D. W. & Barnes, W. L. Effects of dipole orientation and birefringence on the optical emission from thin films. *Opt. Commun.* **183**, 109-121 (2000).

10    Smith, L. H., Wasey, J. A. E., Samuel, I. D. W. & Barnes, W. L. Light out-coupling efficiencies of organic light-emitting diode structures and the effect of photoluminescence quantum yield. *Adv. Funct. Mat.* **15**, 1839-1844 (2005).

11    Kim, J. S., Ho, P. K. H., Greenham, N. C. & Friend, R. H. Electroluminescence emission pattern of organic light-emitting diodes: Implications for device efficiency calculations. *J. Appl. Phys.* **88**, 1073-1081 (2000).

12    Matterson, B. J. *et al.* Increased efficiency and controlled light output from a microstructured light-emitting diode. *Adv. Mater.* **13**, 123-127 (2001).

13    Kim, K.-H. *et al.* Phosphorescent dye-based supramolecules for high-efficiency organic light-emitting diodes. *Nat. Commun.* **5**, 4769 (2014).

14    Kim, S.-Y. *et al.* Organic Light-Emitting Diodes with 30% External Quantum Efficiency Based on a Horizontally Oriented Emitter. *Adv. Funct. Mat.* **23**, 3896-3900 (2013).

15    Schmidt, T. D. *et al.* Evidence for non-isotropic emitter orientation in a red phosphorescent organic light-emitting diode and its implications for determining the emitter's radiative quantum efficiency. *Appl. Phys. Lett.* **99**, 163302 (2011).

16    Frischeisen, J., Yokoyama, D., Adachi, C. & Bruetting, W. Determination of molecular dipole orientation in doped fluorescent organic thin films by photoluminescence measurements. *Appl. Phys. Lett.* **96**, 073302 (2010).

17    van Mensfoort, S. L. M. *et al.* Measuring the light emission profile in organic light-emitting diodes with nanometre spatial resolution. *Nature Phot.* **4**, 329-335 (2010).

18    Graf, A. *et al.* Correlating the transition dipole moment orientation of phosphorescent emitter molecules in OLEDs to basic material properties. *J. Mat. Chem. C*, 10.1039/c1034tc00997e (2014).

19    Shaw, G. B. *et al.* Ultrafast structural rearrangements in the MLCT excited state for copper(I) bis-phenanthrolines in solution. *J. Am. Chem. Soc.* **129**, 2147-2160 (2007).

20    Betzig, E. & Chichester, R. J. Single molecules obersved by near-field scanning optical microscopy. *Science* **262**, 1422-1425 (1993).

21    Guttler, F., Croci, M., Renn, A. & Wild, U. P. Single molecule polarization spectroscopy: Pentacene in p-terphenyl. *Chem. Phys.* **211**, 421-430 (1996).

22    Huser, T., Yan, M. & Rothberg, L. J. Single chain spectroscopy of conformational dependence of conjugated polymer photophysics. *Proc. Natl. Acad. Sci. U. S. A.* **97**, 11187-11191 (2000).

23    Muller, J. G., Lupton, J. M., Feldmann, J., Lemmer, U. & Scherf, U. Ultrafast intramolecular energy transfer in single conjugated polymer chains probed by polarized single chromophore spectroscopy. *Appl. Phys. Lett.* **84**, 1183-1185 (2004).

24    Mirzov, O. *et al.* Polarization Portraits of Single Multichromophoric Systems: Visualizing Conformation and Energy Transfer. *Small* **5**, 1877-1888 (2009).

25    Hofkens, J. *et al.* Probing photophysical processes in individual multichromophoric dendrimers by single-molecule spectroscopy. *J. Am. Chem. Soc.* **122**, 9278-9288 (2000).

26      Vauthey, E. Photoinduced Symmetry-Breaking Charge Separation. *Chem. Phys. Chem.* **13**, 2001-2011 (2012).

27      Mei, E. W., Vinogradov, S. & Hochstrasser, R. M. Direct observation of triplet state emission of single molecules: Single molecule phosphorescence quenching of metalloporphyrin and organometallic complexes by molecular oxygen and their quenching rate distributions. *J. Am. Chem. Soc.* **125**, 13198-13204 (2003).

28      Vacha, M., Koide, Y., Kotani, M. & Sato, H. Single molecule detection and photobleaching study of a phosphorescent dye: organometallic iridium(III) complex. *Chem. Phys. Lett.* **388**, 263-268 (2004).

29      Koide, Y., Takahashi, S. & Vacha, M. Simultaneous two-photon excited fluorescence and one-photon excited phosphorescence from single molecules of an organometallic complex Ir(ppy)(3). *J. Am. Chem. Soc.* **128**, 10990-10991 (2006).

30      Sekiguchi, Y., Habuchi, S. & Vacha, M. Single-Molecule Electroluminescence of a Phosphorescent Organometallic Complex. *Chem. Phys. Chem.* **10**, 1195-1198 (2009).

31      Hu, D. H. & Lu, H. P. Single-molecule triplet-state photon antibunching at room temperature. *J. Phys. Chem. B* **109**, 9861-9864 (2005).

32      Nothaft, M. *et al.* Electrically driven photon antibunching from a single molecule at room temperature. *Nat. Commun.* **3**, 628 (2012).

33      Kimble, H. J., Dagenais, M. & Mandel, L. Photon anti-bunching in resonance fluorescence. *Phys. Rev. Lett.* **39**, 691-695 (1977).

34  Diedrich, F. & Walther, H. Nonclassical radiation of a single stored ion. *Phys. Rev. Lett.* **58**, 203-206 (1987).

35  Basché, T., Moerner, W. E., Orrit, M. & Talon, H. Photon antibunching in the fluorescence of a single dye molecule trapped in a solid. *Phys. Rev. Lett.* **69**, 1516-1519 (1992).

36  Fleury, L., Segura, J. M., Zumofen, G., Hecht, B. & Wild, U. P. Nonclassical photon statistics in single-molecule fluorescence at room temperature. *Phys. Rev. Lett.* **84**, 1148-1151 (2000).

37  Vacha, M. & Kotani, M. Three-dimensional orientation of single molecules observed by far- and near-field fluorescence microscopy. *J. Chem. Phys.* **118**, 5279-5282 (2003).

38  Stangl, T. *et al.* Temporal Switching of Homo-FRET Pathways in Single-Chromophore Dimer Models of pi-Conjugated Polymers. *J. Am. Chem. Soc.* **135**, 78-81 (2013).

39  Aggarwal, A. V. *et al.* Fluctuating exciton localization in giant pi-conjugated spoked-wheel macrocycles. *Nature Chem.* **5**, 964-970 (2013).

Supporting information for "Spontaneous fluctuations of transition dipole moment orientation in OLED triplet emitters"

Florian Steiner, Sebastian Bange, Jan Vogelsang, and John M. Lupton*

Institut für Experimentelle und Angewandte Physik, Universität Regensburg, Universitätsstrasse 31, 93053 Regensburg, Germany

*Sample preparation and experimental setup*: *Tris*(2-phenylisoquinoline)iridium(III) (Ir(piq)$_3$) and poly(methylmethacrylate) (PMMA, $M_n$=46 kDa, PDI=2.1) were purchased from Sigma Aldrich Co. Isolated Ir(piq)$_3$ molecules were embedded in a ~50 nm thick PMMA matrix by dynamically spin-coating at 2,000 r.p.m. from toluene after substrate cleaning and molecule dilution steps as has been described previously [1,2].

Single molecule measurements were conducted using a customized confocal scanning fluorescence microscope based on an Olympus IX71. The excitation light was generated using a c.w. fiber-coupled diode laser (PicoQuant, LDH-D-C-485) at 485 nm and passed through a clean-up filter (AHF Analysentechnik, z485/10), expanded and collimated via a lens system to a beam diameter of about 1 cm and coupled into an oil-immersion objective (Olympus, UPLSAPO 60XO, NA=1.35) through the back port of the microscope, and passed through a dichroic mirror (AHF Analysentechnik, z488RDC) for confocal excitation. The phosphorescence signal passed a bandpass filter (AHF Analysentechnik, 650/200 BrightLine HC), was split by a 50/50 or a polarizing beam splitter and detected by two avalanche photodiodes (APDs, PicoQuant, τ-SPAD-20). Time-tagged photon arrival was recorded by a TCSPC system (PicoQuant, HydraHarp 400) and further analyzed with a LabView routine.

*Second-order cross-correlation analysis:* Using the equation

$$g^2(\tau) = \frac{\langle I_1(t)I_2(t+\tau)\rangle}{\langle I_1(t)\rangle\langle I_2(t)\rangle},$$

with $I_1(t)$ and $I_2(t)$ corresponding to the measured intensities on the detection channel (photodiode) 1 and 2, respectively, the cross-correlation curves were computed. These curves were fitted by using a single-exponential function

$$g^2(\tau) = y_0 + A \cdot e^{-\frac{\tau}{\tau_{ac}}} \text{ with the autocorrelation time } \tau_{ac} = \frac{1}{k_{Exc} + k_T}$$

an offset $y_0$, an amplitude $A$, $k_{Exc}$ as the excitation rate and $k_T$ as the triplet decay rate [3]. Note that at low excitation intensities (small excitation rates), the autocorrelation time $\tau_{ac}$ becomes the phosphorescence lifetime $\tau_{PL} = (k_T)^{-1}$.

*Emission intensity saturation:* The graph in Fig. 1e) is fitted by a saturation function according to ref. [4]

$$I(P_{exc}) = I_\infty \frac{P_{exc}/P_S}{1 + P_{exc}/P_S},$$

with $I_\infty$ the fully saturated photoluminescence emission rate and $P_S$ the saturation excitation intensity.

*Simulations of the linear dichroism histograms for a perfect linear dipole:* Linear dichroism histograms are calculated from a simulated ensemble of $10^5$ individual single-dipole emitters, of which the dipole axes are randomly and uniformly oriented in 3-dimensional space. The linear dichroism in emission is calculated for each dipole by mapping polarization states of light rays in the objective's backplane to the sample plane, effectively taking into account the polarization-scrambling influence of the objective's high numerical aperture of 1.35. To correctly account for the experimental signal and noise levels in photodetection, the individual emitter's brightness and the orientation-independent background signal are taken from Poissonian distributions with mean values of $3.4 \times 10^5$ and $4 \times 10^4$, respectively. To derive the linear dichroism histogram for dipoles that are oriented flat on the surface, a post-selection for out-of-plane angles below 20° is applied to the ensemble of 3D randomly oriented dipoles before calculation of the linear dichroism values.

References:

[1]     Stangl, T. *et al.* Temporal Switching of Homo-FRET Pathways in Single-Chromophore Dimer Models of pi-Conjugated Polymers. *J. Am. Chem. Soc.* **135**, 78-81 (2013).

[2]     Aggarwal, A. V. *et al.* Fluctuating exciton localization in giant pi-conjugated spoked-wheel macrocycles. *Nature Chem.* **5**, 964-970 (2013).

[3]     Kitson, S. C., Jonsson, P., Rarity, J. G., Tapster, P.R. Intensity fluctuation spectroscopy of small numbers of dye molecules in a microcavity. *Phys. Rev. A* **58**, 620-627 (1998).

[4]     Ambrose, W. P., Basché, T., Moerner, W. E. Detection and spectroscopy of single pentacene molecules in a para-terphenyl crystal by means of fluoescence excitation. *J. Chem. Phys.* **95**, 7150-7163 (1991).